%
\magnification=\magstep1
\font\bigbfont=cmbx10 scaled\magstep1
\font\bigifont=cmti10 scaled\magstep1

\vsize = 23.5 truecm
\hsize = 15.5 truecm
\hoffset = .2truein
\baselineskip = 14 truept
\overfullrule = 0pt
\parskip = 3 truept
\def\frac#1#2{{#1\over#2}}


\topinsert
\endinsert
\centerline{
{\bigbfont 
SHOT NOISE IN MESOSCOPIC QUANTUM SYSTEMS}}

\vskip 16 truept

\centerline{\bigifont M. P. Das}
\vskip 4 truept
\centerline{
Department of Theoretical Physics}
\centerline{
Research School of Physical Sciences and Engineering}
\centerline{
The Australian National University}
\centerline{
Canberra ACT 0200, Australia}
\vskip 12 truept

\centerline{\bigifont F. Green}
\vskip 4 truept
\centerline{
Centre for Quantum Computer Technology}
\centerline{
School of Physics}
\centerline{
University of New South Wales}
\centerline{
Sydney, NSW 2052, Australia}

\vskip 1.0 truecm
\centerline{\bf 1. INTRODUCTION}
\vskip 12 truept


The fractional quantum Hall effect (FQHE) was discovered in nearly perfect
two-dimensional electron systems, in the presence of a strong perpendicular
magnetic field
{[1,2]}.
Such a field is sufficiently intense for the magnetic length to become
comparable to the separation between carriers. Furthermore, the
degenerate carriers (originally electrons and, later, holes)
are quasiballistic. Their scattering mean free path is reckoned
in microns.

The physics of current transport is particularly rich and exotic
in the FQHE. Hence there is every reason to expect further
puzzles -- and a few surprises -- in the fluctuation
structure of this strongly correlated quantum Hall fluid.
A particular class of theoretical tenets has come to dominate
most, if not quite all, interpretations of noise measurements.
From one point of view these particular notions could be seen
as extending a relatively standard, semiclassically inspired
treatment of fluctuations to quantum noise in the FQHE.
In their own right, noise experiments for the FQHE are far from prosaic;
they are remarkable {\it tours de force} of experimental skill.

In these strongly quantized conductors, the longitudinal resistance 
displays remarkable oscillations as a function of the magnetic field $B$,
dipping down to zero for finite
intervals of $B$. At the same time, the transverse
(Hall) resistance shows absolutely flat plateaux there.
These features
appear at specific, fractional values of the {\it filling factor}.
The filling factor $\nu$
quantifies the state of the system: it is the ratio of the
electron sheet density to the magnetic field ($B$ can be
expressed in units of flux quanta per unit area,
thus counting the density of magnetic flux lines
that pierce the electron sheet).
When $\nu$ falls below one,
we leave the domain of normal carrier behaviour and enter that of
strong electron-electron correlations.

Below, we identify and critique some of the
now-widespread beliefs about FQHE transport and fluctuations.
We argue that these may not be as solidly grounded in the
microscopics of the problem, as one would expect of a
first-principles transport theory.
This relative lack of well-defined grounding is already
evident in certain semiclassical models of mesoscopic noise
{[3]}.
Similar questions of principle can be posed in more highly
quantum-coherent contexts; after all, essentially the same
transport methodology is claimed to underlie both regimes
(see, for example, Reference {[4]}).
Current noise in the FQHE is a prime example for study.
To set the scene, we first revisit
the basic, low-field current response.

\vskip 28 truept 
\centerline{\bf 2. ESSENTIALS}
\vskip 12 truept


The novel feature of Hall plateaux, observed at highly specific
fractional fillings ($\nu = {1\over 3}, {1\over 5}$,
and so on), was first explained by Laughlin
{[1]}
who devised an elegant many-body wave function to capture the behaviour of
electrons in the FQHE state. It was soon realized
that the elementary excitations of the FQHE must possess
{\it fractional charge} and {\it fractional quantum statistics}.
Following these discoveries, two issues emerged in the
physics of the FQHE.

(1) At low applied voltage the Hall current is carried by the ``edge
states'' located at the physical boundaries of the structure
(see below). To probe this situation, a new transport experiment
was set up
to observe edge-state transport in a Corbino geometry
{[5]}.
Normally, in such a topology, carriers would necessarily
transport current through the two-dimensional bulk.
However, for any quantum Hall
arrangement (both integral and fractional)
the intricate potential landscape of the Landau levels,
as sensed by the quasiparticles, should give rise to edge states.
These would keep their one-dimensional character
while meandering and threading their way right through the
bulk of the Corbino disk.

The upshot of this and other experimental probes was that,
for both integral and fractional quantum Hall situations,
one came to believe that it is {\it only} the edge states
that can channel the observable current.
This leads to an unprecedented situation:
topologically the edge states behave,
to all intents, as strictly one-dimensional conductors
{[2]}
regardless of the two-dimensional nature of
the original carrier states.

Contrary to the wide assumption that the edge states must carry all of
the FQHE current in close neighborhood to the device boundary,
von Klitzing has given a striking experimental counterexample.
Careful Hall-conductance measurements in a Corbino structure
show that, in his words, {\it the current cannot flow exclusively
within a very narrow region close to the edge}
{[5]}.

(2) If a current is flowing, it should be possible to
measure its associated {\it fluctuations} (noise).
Noise exists in both equilibrium and nonequilibrium states of the system.
From the scaling of the nonequilibrium fluctuations,
and specifically those referred to as {\it shot noise},
one should be able to determine the effective
charge quantum of the current-bearing excitations
{[4]}.

\vskip 28 truept 
\centerline{\bf 3. EDGE CHANNELS}
\vskip 12 truept


Quantization of the Hall conductance in the
integral (IQHE) and fractional quantum Hall
effects is strikingly different in each case.
This was pointed out by Beenakker and van Houten
{[6]}.
While a single-particle description (in the necessary
presence of disorder) is invoked to explain the IQHE,
the FQHE arises from many-body interactions alone. Wherever
the electronic Landau levels formed in the region
of the sample boundaries
intersect the Fermi level, they define a set of free
current-carrying orbitals. These are known as the edge states.
The boundaries are, from the carriers' point of view, regions
of very high confining potential.

The response of carriers in the edge channels
should depend directly on the form of the underlying Hamiltonian.
Thus it matters critically
whether the Landau-level scheme is an integral or a fractional one.
In principle, experiments on edge-state conduction in the FQHE
(strictly multi-particle physics) might well be expected to differ
from the IQHE (strictly single-particle physics).
Since, however, the current response
appears rather similar in both, it is tempting
to interpret {\it all} edge-state transport
within a single, implicitly one-body, picture
irrespective of whether the Landau-level filling is integral
or fractional.

The assumption of a common mode for
edge-state transport becomes much less self-evident
in the case of the associated current fluctuations.
Inherently, the latter will probe the many-body
dynamical response even for a system of
strictly {\it non}interacting quantized fermions,
free of short-range exchange-correlation effects.
The manifest and fundamental
distinction between the current, a {\it mean property},
and the noise, its mean-square statistical {\it variance},
goes to the heart of the most basic issues in the FQHE.
While there is a systematic path from the latter to the
former via the fluctuation-dissipation theorem,
there is, of course, no general way of extracting the fluctuations
purely from the mean current-voltage characteristics.
While one may argue whether such issues
have yet to be meaningfully addressed in mesoscopic physics,
there is no doubt that they are basic to
any understanding of the FQHE that claims to be physically complete.

In other words, current fluctuations are exquisitely sensitive to
off-diagonal effects in a correlated system. They reveal much, much
more than is contained in the current response by itself.
The mean current response tells little -- if anything -- of the
{\it uniqueness} of a multi-particle system's internal dynamics;
the richness of its physics lies
in the details of its off-diagonal behaviour.
This more arduous analytical path is not usually the one that
is trodden by popular accounts of FQHE noise.

The edge-state FQHE has been addressed simultaneously
in three different theoretical papers
{[7,8,9]},
which provide three quite different answers to the problem.
Fractional edge channels have also been studied
in the laboratory
by selective probing of the device boundaries
{[7,9]}.
In simple analogy with the IQHE, there has developed a picture
in which FQHE edge channels carry the actual current
via the {\it same} elementary modes that define the
excitations of the FQHE deep in the bulk of the sample.
A recognizably canonical, microscopic derivation of this
intuitive hypothesis has not yet appeared in the literature.

Here one meets a delicate question: Just how is it that
the edge excitations achieve the same internal
configuration that is mandatory for the existence of the
gapped, {\it incompressible} Laughlin quasiparticle states in the bulk?
In our view, the theoretical constructs underpinning the
present consensus on edge-state statistics  -- namely that they
must be isomorphic with the excitations of the bulk FQHE -- do
not enjoy the clarity of reasoning evident
in Laughlin's own theory of the incompressible bulk fractional states.

The appropriate formalism for the edge-state fractional
excitations, and the appropriate experiments that would test it,
are simply not in evidence so far.
Such data as exists may (or may not) be suggestive of
the intuitive consensus on the edge states' fractional inheritance
from the Laughlin bulk. Unfortunately,
we seem still to be far from a compelling proof that the consensus
solution is unique, and that there is no alternative.

\vskip 28 truept 
\centerline{\bf 4. ONE-DIMENSIONAL CONDUCTION}
\vskip 12 truept


Now we focus on a further assumption made for the edge states: the
notion that they constitute a one-dimensional (1D) conductor
very similar to the channel of a perfect 1D quantum wire.
The only difference here is that edge excitations in the FQHE
state are chiral. They move in only one direction around the boundary,
and are physically decoupled from the bulk states (at the same time
that they are supposed to inherit thier fractionality
from the Laughlin bulk).

The edge excitations in the IQHE are equivalent to a 1D
noninteracting electron gas.
The conductance $G$ in this case is understood
in terms of the quantized Landauer-B\"uttiker formula,

$$
G \equiv {I\over V} = 2{e^2\over h}.
\eqno(1)
$$

\noindent
(spin degeneracy yields the factor of two). In realistic cases,
a subsequent transmission factor ${\cal T}$ is to be
multiplied into the right hand side of Eq. (1)
to account for nonideal forward scattering:
if any scattering occurs in the channel, then ${\cal T} < 1$.
For (chiral) edge states in either QHE,
backscattering is automatically suppressed
unless there are transitions that couple opposite edges
(for instance, a constriction in the device).
The Landauer-B\"uttiker formula seems to give a good
description of edge-state transport
in the idealized noninteracting problem
{[10]}.

For real systems, the electron-electron interaction cannot
be ignored either.
In 1D, even the weakest electron-electron interaction causes
low-order perturbation theory to diverge,
suggesting the breakdown of normal
Fermi-liquid theory. One of two possible ways out of this
breakdown is to carry out a careful
dressing of the quasiparticles, either by resumming the most
divergent diagrams or by introducing many-body effects through an effective
exchange-correlation process. This should render the Fermi-liquid
paradigm ``almost'' correct; that is, it becomes
microscopically accurate in some asymptotic sense
{[11]}.

The other scenario is radically different. It proposes a model
of a strongly localized, 1D interacting Fermi system
known as the {\it Luttinger liquid}
{[12]}.
The Luttinger liquid has no relation to the
perturbative Fermi-liquid picture.
In this model, an infinitesimal interaction is enough
to generate a multitude of (bosonic) collective modes.
The excitation spectrum has absolutely no fermionic
quasiparticle contributions in this case.
An electronic state is filled by creation of an
infinite number of bosons. This implies the separation of
spin and charge degrees of freedom. Quite distinct
excitations now carry spin and charge, which no longer
coexist within the same state (as for a normal quasiparticle).
The correlation functions of charge and spin are quite
anomalous in comparison to those of a normal Fermi liquid.
These are but a few of many more differences between normal
electrons and Luttinger excitations.

Measurements of transport in carefully fabricated
quantum wires turn out to show departures from what is understood
for the noninteracting conductance formula, Eq. (1).
The universal conductance jump of $e^2/h$ appears
to be fractionalized.
Therefore it seems natural to invoke the Luttinger-liquid model,
which modifies the Landauer-B\"uttiker
formula by a factor $g$ that multiplies the
right-hand expression of Eq. (1)
{[13]}.
It is to be noted that, formally,
the same renormalized conductance can also arise from Kubo
linear-response theory
{[14]}.
The value $g = 1$ is proper to noninteracting electrons.
Any value other than unity will account
for attractive ($g > 1$) or repulsive ($g < 1$) many-body interactions.

To explain the observed fractional conductance
{[13]}
by the Landauer-B\"uttiker formula, it is imaginatively proposed
that the charge carriers responsible for transport become fractional
and now carry the charge quantum $e^* = ge$, in keeping with
Luttinger-liquid theory. We call the reader's attention to Ref.
{[14]}
for an interpretation of the role of the channel-lead contacts.

In the chiral fractional-charge model,
the linear conductance is again $G \equiv I/V$.
Whichever transport model is adopted, however,
one has to renormalize {\it both} the current $I$ {\it and} the
operative voltage $V$ (immediately across
the source and drain boundaries of the channel).
This holds for {\it all} theories of the
interacting-electron liquid.
It is then easy to show that, be it in
a Fermi liquid or a Luttinger liquid,
the current and voltage are
renormalized to precisely the same extent.
As a result, the conductance given in Eq. (1) stays exactly
as one finds it in a noninteracting system.

In the Luttinger-liquid interpretation, the 1D
quantum wire is purely ballistic,
free from impurities and thus collisionless.
Its resistance can come only from coupling of the channel's
contacts with the macroscopic, and dissipative,
outer source and drain leads.
The disordered leads are normal metallic Fermi liquids, not 1D systems.
Their large number of internal degrees of freedom provides
the excitations that engender the so-called ``contact'' resistance. The
properties of the contacts make no reference to
details of the excitations
that dominate within the channel itself. Indeed,
this conceptual division between the physical channel and
the physical leads is essential to the applicability of
the Landauer-B\"uttiker conductance
{[15]}.
However, the exact role of disorder is yet to be spelled out
microscopically within this picture. One has only
the assertion that the mismatch between densities of states
in channel and leads, is a self-sufficient explanation.

The actual nature of the contact-resistance problem is
far from trivial. It has been considered in some depth by
Fenton
{[16]} among others,
and more recently by Magnus and Schoenmaker
{[17]}.
Even granting the consensus view
on the issue of the densities of states and their mismatch,
this still leaves one to work out a theory
for the intrinsic states of the channel itself.
Let us look at that.

In a provocative theoretical paper
{[18]}
Wen first argued that in the FQHE situation
(at filling $\nu <1$ such that $\nu^{-1}$ is an odd integer)
the edge-current
modes are isomorphic to the Luttinger-liquid modes.
In a quantum-Hall-bar geometry the right- and left-moving
excitations are localized near the top and bottom of the bar,
respectively.
The chemical-potential difference between the top and bottom
edge excitations is identified with the drop
in the Hall potential, normal to the current.
The Hall conductance is obtained as $G = g e^2/h$. Here $g = \nu^{-1}$
is nothing other than the odd-integer fractional filling.
One now considers the situation for inter-edge tunnelling
at a point contact: the tunneling
current between the top and bottom edge states
is claimed to be transported physically by
charges in units of $e^*= e/\nu$.

Another intriguing aspect of the edge-tunneling experiments
is nonlinear scaling of the $I$--$V$ characteristic.
The power-law exponent in the predicted relation $I \sim V^g$
shows a direct proportionality to the inverse of the filling factor,
as expected within the Luttinger picture. As the fractional filling
changes, the exponent should also change, discontinuously.
This is {\it not} what is observed in the laboratory
{[19]}.

At this point it suffices to remark that the theory
of edge construction is yet to be clearly articulated.
Two contrasting points of view are advanced by Beenakker
{[7]}
and MacDonald
{[8]}.
MacDonald argued that, if the edge width is smaller than
the magnetic length, the edge channel would be very sharp,
indeed abrupt enough to support a strictly 1D Luttinger liquid.
One could then have the Laughlin excitations travelling
about the edge and acting as Luttinger excitations.
Against this argument, Beenakker pointed out that
a realistic edge width will be appreciably greater than the
magnetic length; in that case, the edge current must
be carried by normal electrons
rather than fractionalized charge states.
He further argues that, for intermediate edge widths, one might observe
fractional filling factors whose values are different from the bulk.

Experiments on shot noise in FQHE
devices have not been supplemented by
measurements of the actual edge width in their structures.
On the other hand, groups who have measured the quantized conductance
appear to see edge widths that are smooth, rather than abrupt,
on the magnetic-length scale. The edges seem to
support complicated multiple-channel structures,
consistent with Beenakker's account
{[20]}.

\vskip 28 truept 
\centerline{\bf 5. FLUCTUATIONS}
\vskip 12 truept


Shot noise in electronic conduction is a consequence of the
{\it granularity} of the electronic charge. If successive arrival times
for discrete charge carriers at the
collector electrode have a Poissonian distribution,
one obtains the spectral distribtuion for shot noise
$S(0) = 2q{\overline I}$. Here the spectral function
$S(0)$ is the
zero-frequency Fourier transform of the current-current correlation
function; $q$ is the charge quantum of the current carriers
and ${\overline I}$
is the time-averaged current. The above formula for (classical)
shot noise is a famous result,
known in the literature as the Schottky formula.
It is clearly a signature of {\it nonequilibrium}
charge dynamics in the conducting system. Shot noise
exhibits no dependence on temperature, quite
unlike thermally generated noise, whose current
fluctuations at finite temperature $T$,
and low voltage $eV \ll k_{\rm B}T$
(the quasi-equilibrium limit), yield the Johnson-Nyquist
thermal spectrum $S_{\rm JN}(0) = 4Gk_{\rm B}T$.
This typically scales both
with temperature and conductance.

Currently, the dominant theories of electronic conduction and
shot noise in mesoscopic systems all rely on a concept
of {\it independent} quasiparticle motion, namely the
Landauer-B\"uttiker formalism for coherent single-carrier
transmission. Such a description has shown
enormous flexibility in applications to transport
in multifarious physical systems
{[4]}.
At the same time, it pays little heed to any role for
interactions among the carriers of a mesoscopic system.
In the large majority of mesoscopic studies that have proliferated
during the past decade, many-body
interactions are assumed to play a minor part in the
scheme of things
{[15]}.
It is widely supposed that -- if they matter at all -- the
interactions can be included
as a kind of perturbative afterthought.

Below we look, qualitatively, at just a few of
the reasons why consideration of the interactions is by no
means an optional extra. Indeed, many-body interactions
lie at the very heart of fluctuation kinetics.
Any model failing to account for them properly,
especially in degenerate Fermi liquids,
risks serious inconsistency with the conservation laws
and the sum rules
{[3]}.
Without a guarantee of microscopic conservation,
there can be no assurance that one's predictions
make physical sense.

\vskip 28 truept 
\centerline{\bf 6. SHOT-NOISE EXPERIMENTS}
\vskip 12 truept


Two experimental groups, Saminadayar {\it et al}.
{[21]}
at Saclay and de Picciotto {\it et al}.
{[22]}
at the Weizmann Institute, 
have reported the direct observation of fractional
charges in FQHE systems. These works measured
the shot-noise spectral function for the current.
Measurements were done on the edge states of a
high-mobility two-dimensional electron gas, within a
quantum point contact. This geometry was created by controlling
the bias voltage on a split gate, constricting the 2DEG locally
and bringing the edge states close enough for tunnelling to occur.
Filling factors, either ${1\over 3}$ or ${1\over 5}$, were
fixed at will by modulating
the areal densities of the electrons and the magnetic flux.

In this way shot noise could be
measured at very low frequencies and
at very low temperatures (in the range
of  mK).  The current cross-correlation (essentially shot noise)
was found to be proportional to  the
back-scattered current $I_{\rm B} = I_0 - {\overline I}$, where
${\overline I}$ is
the mean transmitted current and $I_0 = (e^*/h) eV$ is the (chiral)
current coming into the quantum point contact. Most importantly,
$e^*$ is the elementary quasiparticle charge
(${e/3}$, ${e/5}$, etc.).
The results from  both Saclay and Weizmann
generally show that the shot  noise scales
nicely  with the back-scattered current.
The best-fit slope scales with $e^*$, giving
evidence for fractional charge quantization.

On closer comparison of the two experimental accounts,
one finds some significant differences in data interpretation.
Here we recall that, according to Landauer-B\"uttiker theory,
the Schottky formula requires correction by the
Fano factor $\gamma$ which, if less than unity, suppresses
the spectral density so that $S(0) = 2\gamma q I$.
In any sort of barrier, such as a quantum point contact,
this suppression  is  governed solely by  the
transmission coefficient ${\cal T} \leq 1$.
For a single conducting channel, the backscattered shot noise
should carry the Fano factor $\gamma = {\cal T}$.

The  data   of  de  Picciotto  {\it et  al.}
{[22]}
was phenomenologically  fitted  by  choosing a
certain  value of ${\cal T}$ to improve agreement with
predictions for backscattering;
then $S(0) \to 2e^*I_{\rm B}{\cal T}$.
Saminadayar {\it et al.}
{[21]},
on the other  hand, fitted their backscattering
data without {\it any}
suppression factor at all, rightly pointing out that
the presence of a suppression
factor would mean that the actual quasiparticle charge must
exceed its fundamental value (in their case, ${e/3}$), to offset
the action of ${\cal T}$ and recover the linear ${e/3}$ slope
as observed by them.
Yet, despite these mutual inconsistencies it is nevertheless asserted that
the fractional charge of the (bulk) Laughlin quasiparticles has
indeed been measured, and in the most direct way possible,
by these shot-noise experiments
{[23]}.

\vskip 28 truept 
\centerline{\bf 7. THEORY}
\vskip 12 truept


Now  we ask  the  question: What  is the  basis for
understanding  the above experiments?
Let us  start with a pure, low-density, interacting 2D
electron gas in a high magnetic field.

The  system  has  degenerate  Landau levels.  As  shown  by Laughlin
{[1]}
the quasiparticles in this  system at fractional filling $\nu = 1/q$
($q$ odd) have fractional charge $e^* = e/q$ as  well as
an energy gap $D = e^2/l$ where $l = \sqrt{\hbar/eB}$ is the
Landau length. The quasiparticle is an electron with $q$ flux quanta
associated with it.

For  a  finite system,  the  flat Landau  levels of the bulk bend
upward in energy owing to the confining
potential. Crossing of the Fermi level with the confining
potential creates the edge  states. For the
IQHE the edge-state quasiparticles are electrons with charge $e$. For
the FQHE, by analogy, the edge-state quasiparticles are taken to be
the Laughlin excitations of the bulk, with charge $e^*$.
Now the edge channels behave like 1D chiral Luttinger liquid.

Theories of quantum noise due to the Luttinger excitations
are available in Refs.
{[24-27]}.
Here,  Luttinger  excitations of the interacting 1D  electrons and
Laughlin  excitations  of  the  bulk FQHE  are  regarded as synonymous.
Kane  and Fisher {[24]} treated  the edge excitations  as a Luttinger
liquid and used a bosonization technique.
There are two tunneling regimes:  strong and weak.

Strong  tunneling leads  to  weak backscattering, and vice  versa. The
experiments with  quantum point  contacts, described in the Section above,
relate  to the weak-tunneling case, with strong backscattering.
The  formula for shot noise is of precisely the same form as
the noninteracting Landauer-B\"uttiker formula.
Chamon {\it et al}.
{[25]}
have discussed the strong and weak tunneling limit of
Luttinger excitations and used a nonequilibrium Keldysh
formalism to obtain nonlinear
current-voltage relationship, as expected for a Luttinger liquid.
Fendley {\it et al}.
{[26]}
have considered a conformal field-theoretic approach to obtain
the shot noise of Luttinger liquids, and have obtained strong-
and weak-tunneling results at finite temperature.

In a more recent paper Sandler {\it et al.}
{[27]}
have considered tunneling between integral states
and various fractional states. Here
the fractionally charged states are not necessarily Laughlin quasiparticles.
Rather, they correspond to solutions of the coupled systems.
Generalizations of these results
to strong- and weak-coupling regimes are made for a quantum-point-contact
geometry.

\vskip 28 truept 
\centerline{\bf 8. SUMMARY AND CONCLUSIONS}
\vskip 12 truept


With a 1D Luttinger liquid as their starting point,
all of these various theories
ultimately  converge to the standard Landauer-B\"uttiker formalism.
This has exhibited great sucess in  understanding a
huge variety of physical systems
{[3]}.
However, for the measurement and interpretation of
shot-noise results,
it is unclear how these approaches, in conjunction with the
Landauer-B\"uttiker picture, manage to retain control of the fluctuation
structure, whose off-diagonal properties manifest the
intrinsically {\it correlated}
nature of such strongly interacting systems.

We believe that the following
points warrant more complete and logically coherent
explanations than are available at this writing.

$\bullet$
For the fractional edge-channel state, commonly considered
as a Luttinger liquid, one needs to have a theory
with truly clear, physically well-formed premises.

$\bullet$
Why should the fractional charge of the Luttinger liquid
be at all the same as the
fractional charge of the bulk Laughlin excitations?

$\bullet$
How do the bulk, gapped, incompressible excitations lose their
gap and their incompressible character in going to the boundary?

$\bullet$
The quantum shot-noise formula has the typical Schottky form,
corrected for the Landauer-B\"uttiker suppression in forward
transmission, $1 - {\cal T}$ (or ${\cal T}$ for backscattering).
True many-body statistics and
correlations are totally missing in this one-body formula.
Recently Isakov {\it et al}.
{[28]}
tried to incorporate exclusion statistics, but
at the cost of even greater conceptual difficulties
that undermine the heuristic state-counting argument
{[4,29]}.

The important issues related to shot noise have been covered
recently in the extensive review by Blanter and B\"uttiker
{[4]}.
While transport and noise are considered mainly in
the context of independent quasiparticles, those authors themselves
state that ``electrons are, however, interacting entities and both the
fluctuations at finite frequencies and the fluctuation properties far
from equilibrium require in general a discussion of the role of the
long range Coulomb interaction.
A quasi-particle picture is no longer sufficient
and collective properties of the electron system come into play.''

To correctly understand both conductance and shot-noise experiments
in a unified way, one has to develop a genuinely nonequilibrium
theory of transport and fluctuations, including the electron correlations
that have so far been neglected (or, at best, averaged over
in an {\it ad hoc} fashion).
The Landauer-B\"uttiker formula is
derivable from Kubo linear-response theory, and this
provides at least one formal link back to the
many-particle density matrix and its inbuilt correlations.
While recognizing the successes of Landauer-B\"uttiker theory,
we feel that the time is now ripe to formulate a practical and correct
{\it many-body} theory for
the complex behavior of correlated mesoscopic systems.

\vskip 28 truept 
\centerline{\bf REFERENCES}
\vskip 12 truept


\item{[1]}
See {\it Nobel Lectures: Fractional Quantum Hall Effect},
Rev. Mod. Phys. {\bf 71}, 4 (1998)

\item{[2]}
T. Chakraborty and P. Pietilainen,
{\it The Fractional Quantum Hall Effect} (Springer, Berlin, 1988).

\item{[3]}
F. Green and M. P. Das, {\it J. Phys: Condens. Matter}
{\bf 12}, 5233 (2000).

\item{[4]}
Ya. M. Blanter and M. B\"uttiker, {\it Phys. Rep.} {\bf 336}, 1 (2000).

\item{[5]}
K. von Klitzing, {\it Physica B} {\bf 184}, 1 (1993).

\item{[6]}
C. W. J. Beenakker and H. van Houten in
{\it Advances in Solid State Physics}, Vol. 44,
edited by H. Ehrenreich and D.Turnbull, (Academic, New York, 1991).

\item{[7]}
C. W. J. Beenakker, {\it Phys. Rev. Lett.} {\bf 64}, 216 (1990).

\item{[8]}
A. H. MacDonald, {\it Phys. Rev. Lett.} {\bf 64}, 220 (1990).

\item{[9]}
A. M. Chang, {\it Solid State Comm.} {\bf 74}, 871 (1990).

\item{[10]}
B. J. van Wees, H. van Houten, C. W. J. Beenakker, J. G. Williamson,
L. P. Kouwenhoven, D. J. van der Marel, and C. T. Foxon,
{\it Phys. Rev. Lett.} {\bf 60}, 848 (1988);
D. A. Wharam, M. Pepper, H. Ahmed, J. E. F. Frost, D. G. Hasko, D. C.
Peacock, D. A. Ritchie, and G. A. C. Jones,
{\it J. Phys. C} {\bf 21}, L209 (1988).

\item{[11]}
See, for example, S. Das Sarma and B. Hu,
{\it Aust. J. Phys.} {\bf 46}, 359 (1993).

\item{[12]}
G. D. Mahan, {\it Many Particle Physics} (Plenum, New York, 1990).

\item{[13]}
A. Yacoby, H. L. St\"ormer, N. S. Wingreen, L. N. Pfeiffer, K. W. Baldwin,
and K. W. West,
{\it Phys. Rev. Lett.} {\bf 77}, 4612 (1996);
K. J. Thomas, J. T. Nicholls, N. J. Appleyard,
M. Y. Simmons, M. Pepper, D. R. Mace, W. R. Tribe,
and D. A. Richie, {\it Phys. Rev. Lett.} {\bf 77}, 135 (1996).

\item{[14]}
C. Kane and M. P. A. Fisher, Phys. Rev. B {\bf 46}, 15233 (1992);
M. P. A. Fisher and L. I. Glazman in
{\it Mesoscopic Electron Transport (NATO ASI Series E)}
ed L. P. Kouwenhoven, G. Sch\"on, and L. L. Sohn
(Kluwer Academic, Dordrecht, 1997).

\item{[15]}
Y. Imry and R. Landauer, {\it Rev. Mod. Phys.} {\bf 71}, S306 (1999).

\item{[16]}
E. W. Fenton, {\it Phys. Rev. B} {\bf 46}, 3754 (1992);
{\it Superlattices and Microstruct.} {\bf 16}, 87 (1994).

\item{[17]}
W. Magnus and W. Schoenmaker, {\it Phys. Rev. B} {\bf 61}, 10883 (2000). 

\item{[18]}
X.-G. Wen, {\it Phys. Rev. Lett.} {\bf 64}, 2206 (1990);
{\it International J. Mod. Phys. B} {\bf 6}, 1711 (1992).

\item{[19]}
M. Grayson, D. C. Tsui, L. N. Pfeiffer, K. W. West, and A. M. Chang
{\it Phys. Rev. Lett.} {\bf 80}, 1062 (1998).

\item{[20]}
M. Ando, A. Endo, S. Katsumoto, and Y. Iye,
{\it Physica B} {\bf 249-251} (1998), pp 426-9.

\item{[21]}
L. Saminadayar, D.C. Glattli,, Y. Jin, and B. Etienne,
{\it Phys. Rev. Lett.} {\bf 79}, 2526 (1997);
D. C. Glattli, V. Rodriguez, H. Perrin, P. Roche, Y. Jin, and B. Etienne,
{\it Physica E} {\bf 6}, 22 (2000).

\item{[22]}
R.  de Picciotto, M. Reznikov,  M Heiblum, V. Umansky,  G. Bumin,
and D. Mahalu,  {\it Nature} {\bf 389}, 162 (1997);
M. Reznikov, R. de Picciotto, T. G. Griffiths, M. Heiblum,
and  V. Umansky, {\it Nature} {\bf 399}, 238 (1999).

\item{[23]}
M. Heiblum and A. Stern, {\it Physics World} {\bf 13}, 37 (2000).

\item{[24]}
C. L. Kane and M. P. A. Fisher,
{\it Phys. Rev. Lett.} {\bf 72}, 724 (1994).

\item{[25]}
P. Fendley, A. W. W. Ludwig, and H. Saleur,
{\it Phys. Rev. Lett.} {\bf 75}, 2196 (1995).

\item{[26]}
C. de C. Chamon, D. E. Freed, and X. G. Wen,
{\it Phys. Rev. B} {\bf 51}, 2363 (1995).

\item{[27]}
N. P. Sandler, C. de C. Chamon, and E. Fradkin,
{\it Phys. Rev. B} {\bf 59}, 12521 (1999).

\item{[28]}
S. B. Isakov, Th. Martin, and S. Ouvry. {\it Phys. Rev. Lett.}
{\bf 83}, 580 (1999).

\item{[29]}
M. P. Das and F. Green, {\it Phys. Rev. Lett.} {\bf 85}, 222 (2000).

\end